# GRAVITY:
# The adaptive optics assisted, two object beam combiner for the VLTI

S. Gillessen[1,*], G. Perrin[2], W. Brandner[5], C. Straubmeier[3], F. Eisenhauer[1], S. Rabien[1], A. Eckart[3], P. Lena[2], R. Genzel[1,4], T. Paumard[1], S. Hippler[5]

[1] Max-Planck-Institut für extraterrestrische Physik (MPE)
[2] Observatoire de Paris – site de Meudon
[3] I. Physikalisches Institut, University of Cologne
[4] Department of Physics, University of California, Berkeley
[5] Max-Planck-Institut für Astronomie (MPIA)

## ABSTRACT

We present the adaptive optics assisted, near-infrared VLTI instrument - GRAVITY - for precision narrow-angle astrometry and interferometric phase referenced imaging of faint objects. Precision astrometry and phase-referenced interferometric imaging will realize the most advanced vision of optical/infrared interferometry with the VLT. Our most ambitious science goal is to study motions within a few times the event horizon size of the Galactic Center massive black hole and to test General Relativity in its strong field limit. We define the science reference cases for GRAVITY and derive the top level requirements for GRAVITY. The installation of the instrument at the VLTI is planned for 2012.

## 1. SCIENCE CASES

GRAVITY is a versatile instrument and is able to tackle a wide range of open astrophysical questions. Its main application will be the astrometry and imaging of sources that are red, faint and of complex structure. Therefore, it will be a powerful tool for studying gas motions and stellar orbits around supermassive black holes (SMBH) and active galactic nuclei, for exploring the existence and mass of intermediate mass black holes, for studying stellar masses at both ends of the mass function, for investigating circumstellar disks and jets around young stars, for dynamically detecting and measuring the masses of extrasolar planets, and for studying microlensing events. Here, we concentrate on three science cases: Probing the mass distribution in the central light day of the Galactic Center (GC) by detecting post-Newtonian orbits of stars, probing space-time around the GC SMBH by observing the motion of infrared flares, and measuring the masses of the most massive stars in the Galaxy.

### 1.1. Post-Newtonian orbits

The current best estimates of the mass of the central black hole and the distance to the GC are obtained through orbit fitting of stars in the central arcsecond of the Galaxy, the so-called S-stars. Stellar counts predict that a few faint stars ($17.5 < m_K < 19.5$) should reside even closer to the BH, within the central 100 mas of the Galaxy. These stars have orbital periods of the order of one year, periapse distances of about 1 mas ≈ 100 $R_S$, and travel at relativistic velocities of ~ 10 % of the speed of light during their periapse passages. The repeated interferometric imaging of these stars will allow probing the mass distribution within ~100 Rs to test for the presence of a cluster of dark objects (for example stellar mass black holes). Such a dark cluster might have formed through mass segregation and will only be observable through its gravitational field. It is predicted by theories that explain the growth of SMBH in terms of successive merger events of intermediate mass black holes. Testing if such a cluster is present in the GC therefore directly addresses the long standing puzzle of how SMBH obtained their mass. In addition we will probe relativistic effects in the orbits, in particular the prograde periapse shift (Figure 1).

---
[*] ste@mpe.mpg.de

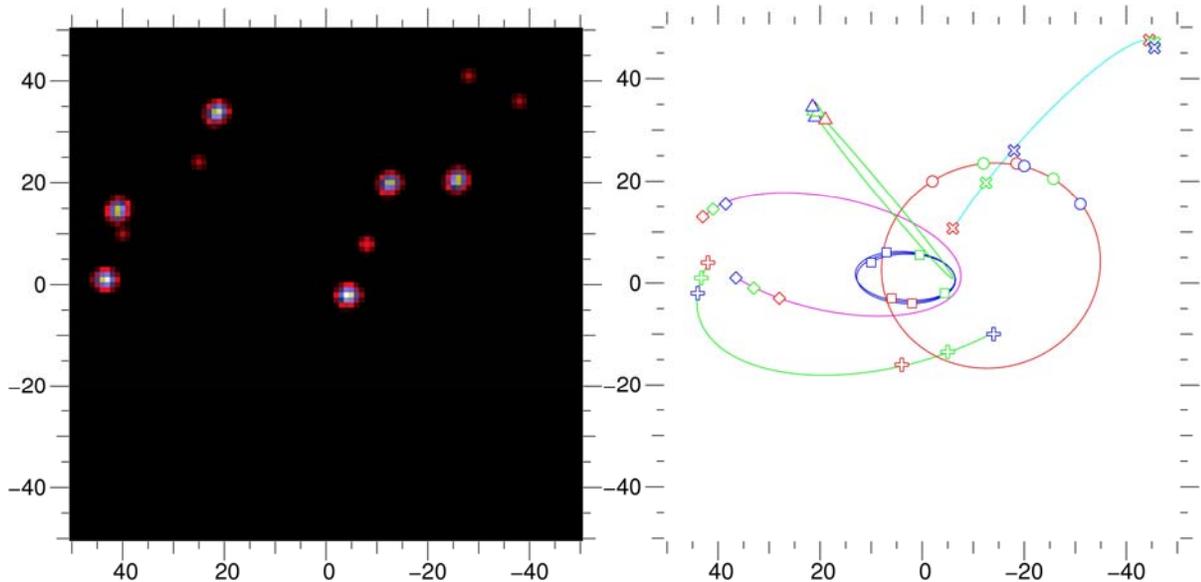

Figure 1: Left: Simulated image of the innermost 100 mas of the Galaxy as GRAVITY will be able to see it. The simulation assumes this region of the cusp contains 6 stars with a maximum brightness difference of 1 mag and 9 hours of observations with the four 8m telescopes of the VLT. The real configuration of the VLT has been used. Noise has been added to the visibilities and phases, and the image has been synthesized and deconvolved using the standard CLEAN algorithm. All 6 stars are recovered. Right: 15 months of astrometry using the image synthesis technique from the left panel allow tracing the orbits of the 6 stars. Several stars have completed at least one orbit, and the relativistic periapse shift is already significant for at least one star (axes in mas).

### 1.2. Probing space-time with infrared flares from SgrA*

Sgr A* exhibits outbursts of infrared, X-ray, and submillimeter emission typically a few times a day. These "flare" events last for about 1 h, and their light-curves show significant variations on a typical timescale of 15 – 20 min in the IR. Given the typical rise time of the substructures in the light-curve, they must come from regions smaller than about 10 light-minutes, or ≈ 17 $R_S$ ≈ 150 µas. Recent evidence from multi-wavelength studies and infrared polarimetry (Eckart et al. 2006, Gillessen et al. 2006) strongly suggests that the emitting region in the NIR consists of a compact (smaller than the Schwarzschild radius) spot of hot gas emitting synchrotron radiation. Such a spot cannot remain static in the potential well of the BH, but its velocity must be comparable to the Keplerian circular velocity $v \approx (r/R_S)^{-1/2} \times 10$ µas min$^{-1}$ ≈ 1 µas min$^{-1}$. Therefore, measuring the 2D astrometry of flares with 10 µas accuracy and a time resolution of a few minutes will not only allow the determination of the location of the flares with respect to the SMBH, but also their proper motions.

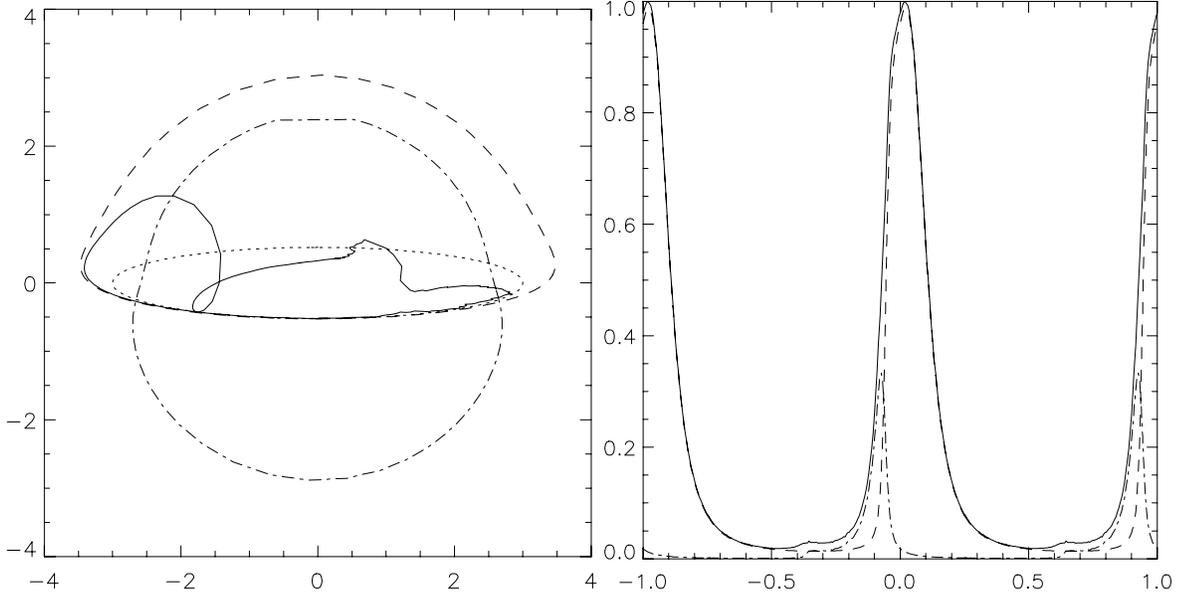

Figure 2: Photo-center wobbling (left) and light curve of a hot spot on the innermost stable orbit around a Schwarzschild BH, at an inclination of 80° as derived from ray tracing in a Schwarzschild metric. Dotted curve: true path of the hot spot; dashed curves: apparent path and light curve of the primary image; dash-dotted curves: same for secondary image; solid curves: path of centroid and integrated light curve. Axes of left panel in Schwarzschild radii, roughly equal to the astrometric accuracy of 10 μas. Abscissa axis of right panel in cycles. The loop in the centroid's track is due to the secondary image. The overall motion can be detected at good significance at the anticipated accuracy of GRAVITY. Details can be obtained by analyzing several flares simultaneously (Paumard et al. 2006).

Additionally, the flares are a dynamical probe in the potential well of Sgr A* at a few $R_S$. If the observed ~ 17 min quasi-periodicity seen in integrated light in several of the infrared flares is caused by a hot spot near the last stable orbit of a Kerr BH, it is possible to deduce the BH's spin parameter ($J/J_{max}$ ~ 0.52, Genzel et al. 2003). If a simple orbiting hot spot model is indeed applicable (to some of the flares), then 10 μas-accuracy astrometry may be a clean enough dynamical probe of the space-time down to the photon-sphere of the BH (Broderick & Loeb 2005, 2006; Paumard et al. 2005, 2006; Figure 2). In this instance, GRAVITY will test General Relativity in its strong field limit. This ambitious experiment requires the simultaneous, multi-baseline, narrow-angle astrometric capability of GRAVITY.

### 1.3. Masses of the most massive stars

There still exists an uncertainty of up to a factor of two in the mass estimates from photometry and spectroscopy for the most massive stars. Comparing spectra with atmospheric models yields upper mass limits in the range of 60 $M_\odot$, whereas evolutionary tracks and observed luminosity suggest a mass of up to 120 $M_\odot$ for stars of spectral type O2V and O3V. Clearly, dynamical mass estimates for early O-type main-sequence stars are required. Luckily, quite a number of spectroscopic binary O-stars are known in the cores of Galactic starburst clusters like the Arches, the Quintuplet and NGC 3603, or extragalactic starbursts like 30 Doradus. With a nominal resolution of 4 mas at a wavelength of 2 μm, GRAVITY could resolve some of the longer period spectroscopic binaries, and monitor the astrometric motion of the photo-center for the shorter period, closer binaries with a precision of 10 μas. Astrometric orbits for these deeply embedded binary stars will hence directly yield dynamical mass estimates. The unique narrow angle astrometry mode of GRAVITY is ideal for these dynamical studies in crowded regions. Several objects (e.g. the Arches, the Quintuplet) cannot be observed without the infrared wavefront sensing provided by GRAVITY (Figure 3).

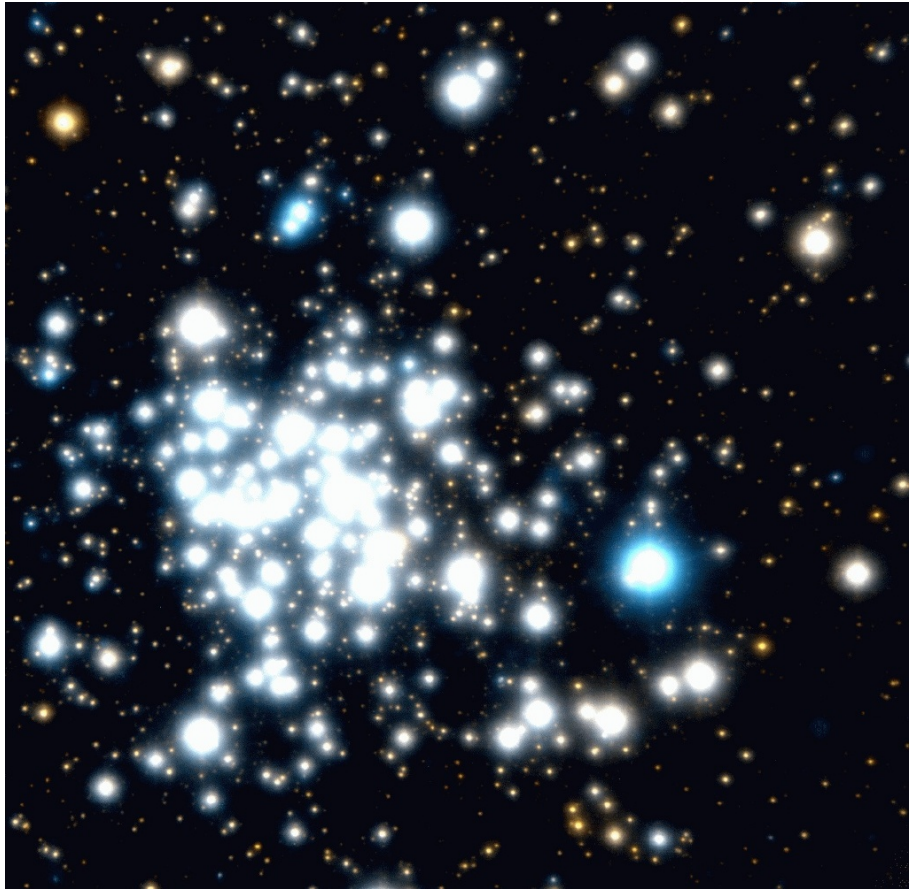

Figure 3: Composite color H- and K-band image of the Arches starburst cluster in the Galactic Center region with a resolution of approx. 60 mas, i.e. about 500 AU (Stolte et al. 2005). GRAVITY will resolve binaries with separations of less than 30 AU, and monitor the photo-center movements of still closer binaries, resulting in dynamical mass estimates for the most massive stars. Monitoring of the dynamics of the central cluster stars should also answer the question of whether or not starburst clusters house intermediate mass black holes in their centers.

## 2. TOP LEVEL REQUIREMENTS

Following the requirements from the science cases, GRAVITY is designed for:

- infrared wavefront sensing down to $m_K > 10$;
- internal fringe tracking down to $m_K > 10$;
- multiple baseline narrow angle astrometry with 10 μas accuracy for operation with the VLTI 8m telescopes (maybe also replace "UT" elsewhere);
- interferometric imaging of faint objects with $m_K > 19$ in 1 hour observing time.

The expected instrument's performances are summarized in table 1.

| Infrared wavefront sensor | Limiting magnitude | Number of sub-apertures |
|---|---|---|
| | $m_K > 10$ | ≈ 60 (Strehl ratio ≈ 50 % @ 2.2 μm using the 60-actuator MACAO bimorph deformable mirrors) |
| **Fringe tracking** | **Limiting magnitude** | **Maximum distance from science object** |
| | $m_K > 10$ | 2 arcsec (VLTI FoV) |
| **Narrow-angle astrometry** | **Accuracy** | **Maximum separation** |
| | ≈ 10 μas for use with the VLT 8m telescopes (a factor 10 better than previous instruments) | 2 arcsec (VLTI FoV) |
| **Interferometric imaging** | **Limiting magnitude** | **Accuracy** |
| | When fringe tracking $m_K > 19.5$ in 1 hour | < 1% in visibility < 1° in phase |
| | In short exposure mode $m_K > 12$ | |

Table 1: Summary of the GRAVITY performance: For UT operation, GRAVITY outperforms existing first-generation instruments in narrow-angle astrometry by a factor 10. GRAVITY provides simultaneous astrometry for up to 6 baselines. GRAVITY is the only instrument that takes advantage of the unique 2" field of view of the VLTI, providing internal fringe tracking for deep interferometric imaging.

## 3. INSTRUMENT CONCEPT

GRAVITY will be developed as an end-to-end general use facility at the VLTI (Glindemann et al. 2004) and will be operated within the ESO data flow system. The instrument will be installed in the VLTI laboratory. In its baseline design, GRAVITY will have infrared wavefront sensors and command the deformable mirrors already installed in the telescopes. The entire beam combiner including field selector, injection optics, fiber optics, integrated optics, camera and detector is enclosed in a cryostat. This not only drastically improves the stability and cleanliness of the instrument, but also allows for optimum baffling and suppression of thermal background. A metrology system measures the OPD between the two beams. The light from the eight fibers (2 objects times 4 telescopes) is then fed to an integrated optics beam combiner. This technology guarantees optimum throughput and stability at little complexity. Various detector options are presently investigated, with focus on large pixels, high quantum-efficiency, and low noise.

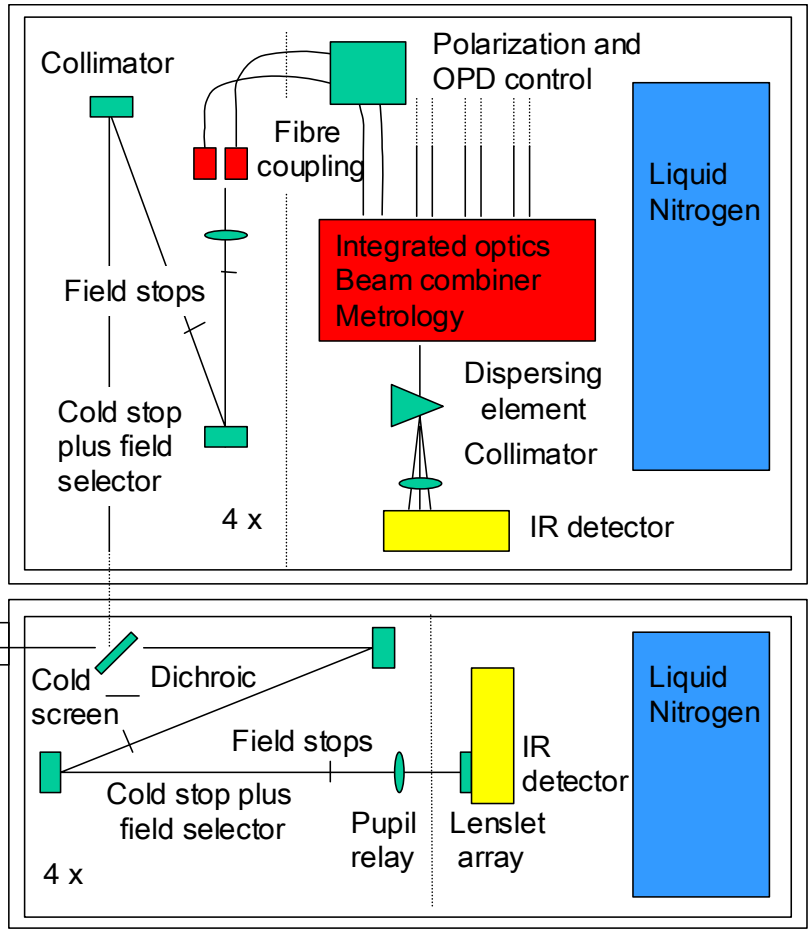

Figure 4: Instrument concept: GRAVITY is made from two modules: infrared wavefront sensor (bottom) and beam combiner / fringe-tracker (top). The instrument is installed in the VLTI laboratory. Therefore, the wavefront sensor also corrects the wavefront errors and seeing introduced in the VLTI beam relay. The wavefront sensor can also be used for observations with other VLTI instruments. The baseline concept makes use of the deformable mirrors already installed at the UTs. The beam combiner is fully enclosed in a cryostat for optimum stability, cleanliness and suppression of thermal background. Its field selector allows picking two objects in the 2" field of view of the VLTI science beam. The light is fed to single-mode fibers for beam cleaning and optical path difference control. An integrated optics beam combiner guarantees optimum efficiency. One of the objects can be used for internal fringe tracking, i.e. GRAVITY will not only provide closure phases, but the differential OPDs for up to 6 baselines (4 telescopes). Taking advantage of the significantly reduced non-common light path and its internal metrology, GRAVITY allows narrow angle astrometry with an accuracy of < 10 µas, a factor of ten better compared to existing first-generation instruments operating with the 8m telescopes.

### 3.1 Adaptive optics

GRAVITY provides four infrared wavefront sensors for the $1 - 2.5$ µm wavelength range. These IR wavefront sensors are used when no adequate ($m_V < 15$) visible guide star is available for the visible wavefront sensing. The GRAVITY IR wavefront sensors make the VLTI an even more unique facility worldwide. GRAVITY corrects wavefront errors and seeing introduced in the VLTI beam relay. Hence, the wavefront sensors are installed in the VLTI laboratory, providing optimum sensing and correction of atmospheric wavefront errors and seeing introduced in the VLTI delay lines and beam relay. Compared to the MACAO (Multi Application Curvature Adaptive Optics, Arsenault et al. 2003) wavefront sensors, which are already installed in the telescopes, the non-common light path is of the order 100 m shorter, and has on the order of ten fewer non-common path mirrors. The GRAVITY wavefront sensor can be operated either on a guide star within the 2" field of view of the VLTI beam, or can be fed with the light from an off-axis guide star picked with star-separators that are already installed at the telescopes. The limiting magnitude for adequate adaptive optics correction is approx. $m_K \approx 10$. In its baseline for operation with the UTs, the GRAVITY wavefront sensors will command the MACAO deformable mirrors already installed at the Coudé focus. The performance will be nearly diffraction limited with a Strehl ratio $\approx 50\%$ and coherent energy $\approx 50\%$ in K-band.

### 3.2. Beam combiner

GRAVITY provides simultaneous interferometry of two objects in a 2" field of view for up to four telescopes. The large 2" field of view of the VLTI delay line is worldwide unique on a 140 m baseline, and no other VLTI instrument is taking advantage of that outstanding capability. GRAVITY delivers narrow angle astrometry with an accuracy of < 10 µas, a factor ten better than previous instruments operating with the 8m telescopes. Compared to the according mode of the PRIMA (one of the first generation instruments that is optimised for planet search and uses the auxiliary telescopes, Delplancke et al. 2000) dual feed facility, GRAVITY does not split the light of the stars before the delay line, but takes advantage of the significantly smaller non-common light path in a single beam for optimum phase accuracy, resulting in significantly smaller systematic errors in the phase measurement. The feasibility of 10 µas narrow angle astrometry, when working on a single beam and taking advantage of fringe tracking, has already been demonstrated at the Palomar Testbed Interferometer (Lane & Muterspaugh 2004).

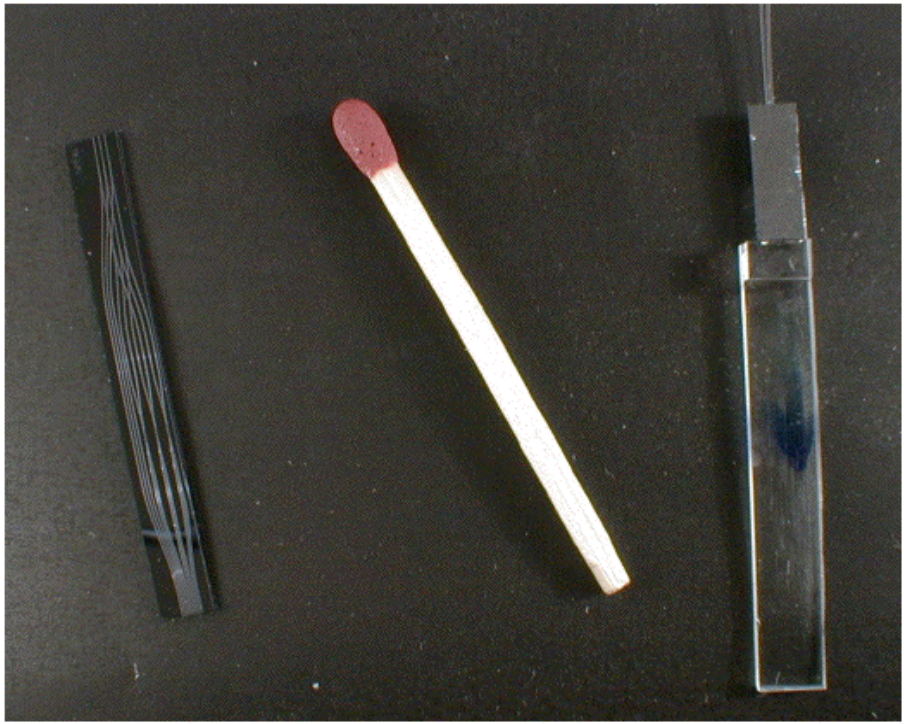

Figure 5: Example of a near infrared three-telescope integrated optics beam combiner made by LETI (**http://www-laog.obs.ujf-grenoble.fr/Recherche/RechercheInstrumentale/hra/ionic/index.html**)

With GRAVITY simultaneous astrometry for multiple baselines becomes possible. For comparison, the PRIMA facility and the Keck interferometer provide astrometry for a single baseline only. The instrument will provide internal fringe tracking for up to four telescopes. The brighter of the two objects within the 2" field of view will serve as a reference star for fringe tracking. Optical path fluctuations (introduced by the atmosphere and the beam relay) will be measured in real-time between the up to four telescopes. These path length fluctuations will be compensated by commanding the delay lines already installed at the VLTI, or with an internal active device if necessary. The minimum brightness required for fringe tracking with GRAVITY is comparable to that of the already existing fringe sensors, and is $m_K \approx 10$. Compared to the PRIMA facility, which splits the light of the phase reference star and the science object before feeding the delay lines, the GRAVITY fringe sensing will benefit from the significantly shorter non-common light path (which is of the order 100 m longer and 10 more mirrors for PRIMA).

## 3.3. Metrology

GRAVITY will be equipped with its own path length measurement and compensation system. Measuring the angular separation of two objects with an accuracy of 10 µas sets high demands on the accuracy to which the instrumental optical path difference (OPD) has to be known. With the largest baselines available within the VLTI, typical errors allowed are on the order of 5 nm. For a two-baseline interferometric setup such as in PRIMA, the standard solution for measuring the differential OPD is a dual wavelength super-heterodyne laser metrology (Leveque et al. 2003).

We will use another concept: backward propagation of laser beams through the entire VLTI. The concept is illustrated in Figure 6. Injecting the laser light into the science fibers will take place in the temperature-stabilized part of the instrument and before any active phase control. This can either be done by wavelength dispersive elements – as dichroic fiber splitters – or the injection takes place within the beam combination unit. When overlapping in a pupil plane, the two laser wavefronts create an interference pattern. Using phase-shifting technologies, or adequate Fourier filtering methods, the relative phase difference between the beams can be measured with high accuracy. For detecting the fringe pattern we intend to use commercially available cameras. With the high number of photons within the laser beams we expect that even the scattered light from the VLT secondary mirror can be used for detection. The laser fringe spacing and phase movement can be measured at high frequency, delivering the OPD difference, i.e. the projected angular separation of the fibers on the sky. The laser wavelength will be close to the observing wavelength, in order to minimize dispersion effects and incomplete mode matching in the fibers.

The main advantage of that concept is that the GRAVITY metrology will cover the full beam. Besides that it is important to notice that the GRAVITY metrology will include the full coude path, which includes the deformable mirror. This will allow GRAVITY to measure and compensate for any instrumentally induced path differences.

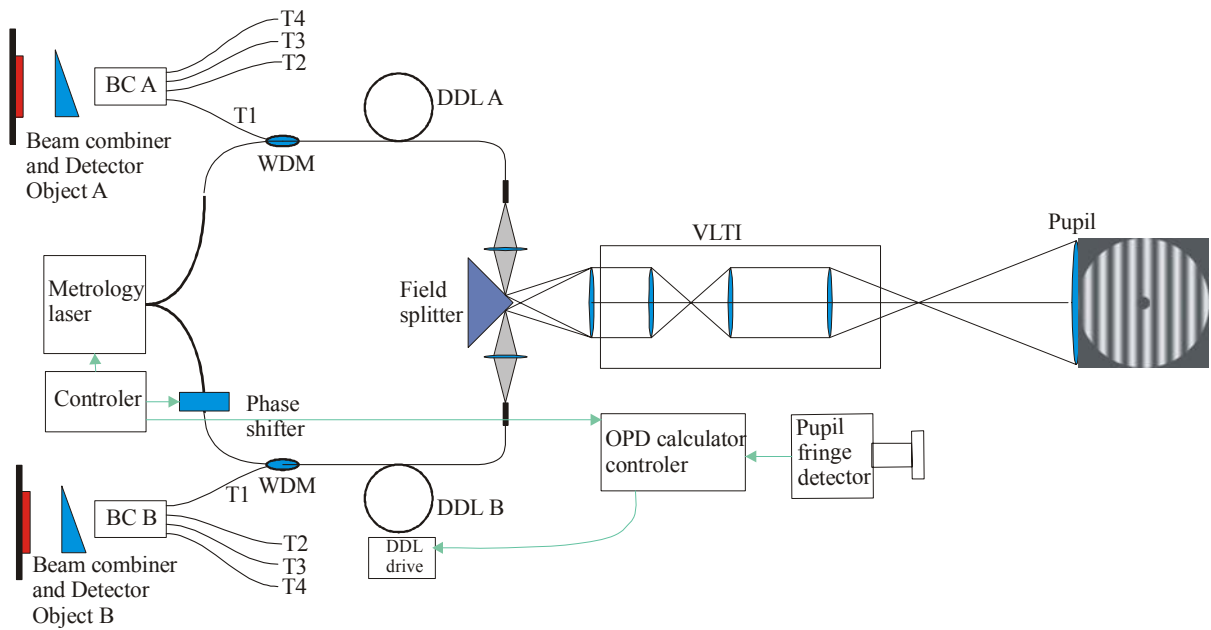

Figure 6: Concept of a differential OPD measurement and control with pupil laser fringe sensing. The metrology laser is injected into the science beam fibers close to the beam combiner, and propagates the whole way backwards through the instrument internal delay lines (DDL), the field selector and the VLTI beam relay. The interference pattern is then detected with an infrared camera installed at the telescope, providing the differential OPD between the two beams. In contrast to the PRIMA metrology system, the laser beams cover the full aperture to minimize systematic errors.

## 4. Conclusions

The VLTI is the only array of 8m class telescopes that explicitly included interferometry in its design and implementation. No other array is equipped with a comparable infrastructure. The VLTI, with its four 8m telescopes, is the only interferometer to allow direct imaging at high sensitivity and high image quality. The two 10m Keck telescopes can only perform astrometry and phase referenced imaging on a single baseline. The VLTI is also the only array of its class offering a large 2" field of view (40 times the diffraction limit of the 8m Unit Telescopes in K-band). By combining GRAVITY with the VLTI, it will be possible to use to full capacity all these basic unique features. VLTI+GRAVITY will thus be a unique facility worldwide for many years to come.